# Experimental study on Modified Linear Quadratic Gaussian Control for Adaptive Optics

Qiang Fu,[a,b,c,d] Jörg-Uwe Pott,[a] Feng Shen,[b,c] Changhui Rao[b,c], Xinyang Li[b,c]

[a] *Max-Planck-Institut für Astronomie, Königstuhl 17, D-69117 Heidelberg, Germany*
[b] *The laboratory on Adaptive Optics, Institute of Optics and Electronics, Chinese Academy of Science, Chengdu, 610209, China*
[c] *The key laboratory on Adaptive Optics, Chinese Academy of Sciences, Chengdu, 610209, China*
[d] *Graduate School of Chinese Academy of Science, Beijing, 100084, China*
*Corresponding author: fqqq1234@163.com

Classical adaptive optics (AO) usually uses the simple but efficient control algorithm like proportional-integral (PI) to achieve high resolution imaging. In the point of view of the minimum variance controller, it is not optimal. So far, Linear Quadratic Gaussian Control (LQG) is an interesting control algorithm with global optimum. However it still encounters one unexpected problem in practice, leading to the divergence of control in AO. Hence, the Modified LQG (MLQG) is proposed in this paper and analyzed explicitly. The test in the lab shows strong stability and high precision compared to the classical control.

**1: Introduction**

Due to the atmospheric turbulence, the ground-based telescopes can not achieve the expected diffraction limit. Adaptive optics (AO) has become a main technique to improve the performance of these telescopes [13]. The basic feedback control loop is usually adopted by the AO system since the time delay is of great significance. The simple control algorithm Proportional plus Integral (PI) control law is still the mainstream of the AO systems. However, the PI control is not the optimal control algorithm and limitations of AO application such as Multi Conjugated AO (MCAO) should be drawn attention [12, 26]. The efficient AO control algorithm is developed by means of optimizing gain of the control parameter, such as optimized modal gain integrator (OMGI) [7,10,11] which is not the optimal algorithm. Linear Quadratic Gaussian (LQG) introduced in 1993 in AO was considered to be the promising control algorithm in AO and has been developed by many research groups [2-6, 8-9, 12, 14-22, 24-30] in the world. It is built on the base of the minimum mean-square error estimator. The interior merits such as needlessly precise knowledge of the priors and the precise control [17] arouse interests of AO researchers.

Although the LQG control has been successfully applied in many aspects such as vibration filtering [25], MCAO [17, 26], and extreme AO (XAO) [31-33], there are also some restrictions. Stability of atmosphere turbulence and measurement noises are two preconditions of the LQG applied in AO. Usually, the measurement noises, like the noise of devices is stable with the same variance, but the atmospheric turbulence is not stable and even varies abruptly depending on the weather conditions. Moreover, we can only be informed of the statistical property of atmosphere. Up to now, the divergence of the control has been to some extent central to some applications such as paper machine headbox [35], GPS and lower Cost INS sensors [36] and Accuracy Improvement of a Laser Interferometer [36], etc. So far, only the weak turbulence or weak wavefront error has been introduced in the lab and corrected by LQG method in AO. However, the divergence is probable and is rarely taken into account. The LQG control in AO is proved to be more and more fruitful, although the divergence issue is still a concern.

In this paper, we present the unstable control of LQG algorithm and determine influential factors of divergence via a set of experiments that is conducted on a classical AO system. The modified LQG control is then proposed and is inspected in the experiment. Comparison with the PI control law presents the advantage of the optimal control. The paper is organized as follows. Section 2 reviews the classical PI control and the state-of-art LQG control. The modified LQG is described in the third section and corresponding results are showed in Section 4 compared to PI control.

**2 Review of the LQG control methods**

Regarding to the control of the state space equation, the object of AO control is the multi-input and multi-output system. If it is considered on the model of classical control, the performance has to be analyzed from one by one channel, to get optimally global control. Thus, the control is described below by state equation. Since classical plus Integral (PI) control is mature and still popular in most of the AO systems, we only compare it to LQG control.

The general system state space model is usually described as linear time-invariant case in the form:

$$x_{k+1} = Ax_k + Bu_k + v_k \quad (1)$$
$$z_k = Hx_k + w_k \quad (2)$$

Eq.(1) is called as system object equation, and Eq. (2) is the measurement equation. Coefficients $A$, $B$ and $H$ are matrix of appropriate dimensions. $v_k$ and $w_k$ are decorrelated zero-mean white Gaussian noise with covariance matrix $Q$ and $R$ respectively. Actually, any existing linear AO controller is equivalent to an observer. It is supposed that the turbulent phase dynamics is

$$\varphi_{k+1} = A_s \varphi_k + v_k \quad (3).$$

$A_s$ is the component of the matrix $A$. This assumption is proposed by Brice Le Roux etc [17] and works very well in the optimal control of MCAO. Therefore, the state vector can be decomposed by

$$x_k = \begin{pmatrix} \varphi_k \\ u_k \end{pmatrix} \quad (4)$$

Therein, $\varphi_k$ is the measurement and $u_k$ is the control vector. Consider a simple integrator control with recurrence equation

$$u_{k+1} = u_k + GS_k \quad (5)$$

$G$ is integral gain. $S_k$ is the noise compared to the real noise $v_k$ in Eq.(3). Eq.(5) is similar to Eq.(3)., hence if the best prediction $\varphi_{k+1}$ is obtained, the control $u_{k+1}$ can be also given by

$$u_{k+1} = P\phi_{k+1/k} \quad (6)$$

$P$ can be calculated by generalized inverse

$$P = (N^T N)^{-1} N^T \quad (7)$$

since $\phi_{cor}^k = Nu_k$. The target of the control is to get the best gain $G$ in Eq.(5) to optimize $u_{k+1}$. We can usually obtain the optimum $G$ by manual adjusting. $G$ is also usually a constant and should be tested for each PI controller. Using anything else than gain $G$ leads therefore implacable to a sub-optimal control with respect to the minimum variance criterion.

**A. Prediction control way-Linear Quadratic Gaussian control**

A classical prediction equation of LQG control can be described as follow.

$$x_{k+1/k} = Ax_{k/k-1} + Bu_k + AL(y_k - H_k \bar{y}_k) \quad (8)$$

Eq.(8) is the close-loop state equation with the noise $v_k = y_k - M\bar{y}_k$ in Eq.(1). $y_k$ is the measured output state and $\bar{y}_k$ is the output predictor. In our case, the measured state is the slopes of the wavefront and the input state is the Zernike mode. Here we want to look for an optimal observation gain $L$ to achieve the optimal prediction $x_{k+1/k}$. It is equal to:

$$L = C_k H_k^T (H_k C_k H_k^T + C_w)^{-1} \quad (9)$$

$C_k$ is estimated as error covariance of the predictor state matrix $x_{k+1/k}$. $C_w$ is the noise covariance of the measurement output. $C_v$ is the error covariance matrix of the estimated state and can be presented as

$$C_v = C_x - A^T C_x A \quad (10)$$

where $C_x$ is the covariance matrix of the input state $X_k$. $C_k$ can be given by solving the Ricatti equation:

$$C_{k+1/k} = AC_{k/k-1}A^T + C_v - AC_{k/k-1}H_k^T (H_k C_k H_k^T + C_w)^{-1} H_k C_k A^T \quad (11)$$

At the first sight, this equation may plunge us into despair, as it seems to be incompatible with real-time constraints, but it does not depend on the measurement. It can be calculated off-line, and even be replaced by its constant asymptotic solution $L_{as}$ in Eq.(9). This is the prediction process separated from whole control process by separation theory in Least Square sense.

According to the separation theory, optimal estimator and optimal control can be separated. We can get the optimal predictor by Kalman filter and get the optimal control by the following equation.

$$u_k = P\phi_{k+1/k}^{tur} \quad (12)$$

$P$ is the projector of the phase on the DM basis and is equal to the operator in Ref.[17]. The discrete-time optimality criterion is

$$J = \lim_{n \to +\infty} \frac{1}{n} (\sum_{k=1}^{n} \frac{1}{T} \int_{(k-1)T}^{kT} |\phi - \phi_k|^2 ds) + \lim_{n \to +\infty} \frac{1}{n} \sum_{k=1}^{n} |\phi_k - Nu_k|^2 = J_{r,1} + J_{r,2} \quad (13)$$

The minimizing $J$ is equivalent to minimize $J_{r,2}$ since only the second term depends on the control vector $u_k$. We could get the best control $u_k$ via the optimal discrete-time control law.

As stated above, LQG linear control theory is companied with the theoretical object state which should be built before the control. There are two different types of turbulence model [17]. The first one is the phase generation model, which is used to generate the time series of the turbulent phase. The second type is prior model. The classical atmosphere model is built as the second one which is the first order Auto-Regression (AR) turbulence model. In the first order AR model, matrix $A$ is diagonal and its elements are adjusted according to the temporal evolution of the turbulence.

The characteristic time of evolution of the AR generated turbulence is defined as the correlation time at $1/e$ where $e$ is the natural exponent. The decorrelation of the first order AR turbulence is exponential, which corresponds to a Power Spectral Density (PSD) that is a constant before a cutoff frequency $f_c$ and then decreases with $f^{-2}$ law. AR turbulence model contains more energy at high temporal frequencies. A higher-order AR model may produce PSD closer to a Taylor model. However, the higher order may increase the computation burden and thus decrease the control stability.

## 3. Modified LQG control

The Kalman filter usually diverges when it is applied to correct the aberration in the atmospheric turbulence since the turbulence is always not stable and the covariance of wavefront also fluctuates. Therefore, accurate state-space model is almost impossible to be obtained. The normal idea is to update the statistical state of Eq.(8)-Eq.(11) every several minutes in LQG control. The other idea is to suppress the divergence in real time by some restrictions. In this paper, we only consider the second way to suppress the divergence since apparently the first one is not able to suppress the error when the measured polluted state due to the noise appears.

Eq.(9) and Eq.(11) are usually used to calculate the gain of the state equation. Once the statistical state equation is obtained, the gain can be calculated offline, since the $C_{k+1/k}$ will converges to the limit after tens steps. We now expand the kalman filter. The initial kalman filter is the iterative and regressive method where the kalman gain is obtained through calculation of the state prediction covariance. Here, the divergence can be suppressed via increasing the gain or reducing the gain in Eq.(12) when the statistical model is built. Then the extra scale $\lambda_k$ is introduced to modify the state in the followed two equations.

$$C_{k/k-1} = \lambda_k A_{k/k-1} C_{k-1} A_{k/k-1}^T + Q_{k-1} \quad (14)$$

$$C_k = (I - L_k H_k) C_{k/k-1} \quad (15)$$

$L$ is a gain same to Eq.(9). Eq.(14) is the covariance matrix of the prediction error and Eq.(15) is the covariance matrix of the filtered error. These two equations are the basically iterative equations in Karlman filter and also the decomposition of the Eq.(11). The key point is to look for the appropriate $\lambda_k$ to modify the predicted state and thus minimize the residual measurement error. In the second section, we review the classical LQG control where one important factor is always ignored, which is that the invariant state equation is not always correct since the atmospheric turbulence is always unstable. Thus two modifications are proposed in the following to restrict $\lambda_k$ in real time.

### A. The reasons of the control divergence of LQG

In the practical LQG control, the divergence happens not only in AO but also in other applications [35,36,37]. However, the reasons why the control diverges are similar.
(1) The rough measurement data due to the unstable setup. This happens in the quite adverse condition like the bad seeing ($r_0$ is only in the scale of several centimeters) when the unexpected fierce wind appears and lasts for seconds. In addition, the control may go out of the scale of the corrector, leading to the controlling crash.
(2) The statistical states especially the covariance of errors deviates from the real states due to the variation of the ambient condition after the control running for several minutes.
(3) The calculation error due to the limit byte size will result in indefinite of the error covariance. This error is only considered on few cases, since nowadays, most of the controlling setup is precision enough.

In the normal LQG control, once the preconditions are obtained via the statistics of the state error and measurement noise, they will be kept unchanging along the whole correction process. If it diverges, the correction has to be stopped and restarted again. Therefore, the two methods are introduced in the following to avoid this defect.

### B. Dealing with the singular value

The drawback of this idea is that we never know the exact instant when the abrupt change of the wavefront happens. The problem appears when the second term of Eq.(13) is destroyed by the abrupt noise. This will lead to the big deviation of the DM control vector. Thus, the LQG diverges and even worse the calculated control vector may be out of the scale range of the DM. This abrupt change is named by the singular value which should be dealt with.

Since the wavefront is predicted by the Zernike modes, the gain $L_k$ could be handled to adjust the control vector $Y_k$ in Eq.(8). A critical criterion should be focused on. It is assumed that the measurement of state $X_k$ is $Z_1, Z_2,…, Z_{k-1}$, prediction value is $X_{k/k-1}$ and the deviation of the predicted state is $d_k$.

$$d_k = Z_k - H_k X_{k/k-1} \quad (16)$$

where the distribution of $d_k$ is the fitting of Gaussian with average equating to 0. The covariance is

$$E[d_k, d_k^T] = H_k C_{k/k-1} H_k^T + R_k \quad (17)$$

Then we judge each element of $Z_k$ depending on

$$|d_k(i)| \leq W \sqrt{H_k C_{k/k-1} H_k^T + R_k(i,i)} \quad (18)$$

If the condition in Eq.(18) is not fulfilled, the $i_{ed}$ line of gain matrix $L_k$ is set by zero and other variables are kept unchanged. $W$ is the scale factor which is usually much bigger than 1 such as 7 or 8. Moreover, the explicit factor depends on the applied environment. This modification deletes the large error state component and of course loses some wavefront information, leading to the degradation of the correction performance. Nevertheless, the residual wavefront error will jump only at the exact moments when the dramatically varying states happen and the divergence is able to be suppressed efficiently.

### C. Modification of the optimal gain

In addition to the singular value in section 3.A, the other noticeable problem, selection of the optimal gain, should be addressed exceptionally. If the statistically priori variances of noise are not stable, the gain $L$ in Eq.(9) are not optimal, even worse leading to the evolution divergence. Therefore, the basic assumption of the previous is difficult to be achieved. An adjustable scale factor on the gain $\lambda_k$ is able to reduce the divergence. This mean we should calculate the variance of the system noise and measurement noise in real time to estimate the best gain and then update the state equation.

The ratio $\lambda_k$ is usually called as forgetting factor since it can increase or decrease the current gain $L$. This means that effect of current variables could be adjusted by way of changing gain. This method is proposed by Qijun Xia etc. in 1994 [35] named by AFKF (adaptive fading Kalman filter). They proposed three methods to solve $\lambda_k$. We will review the simplest one and also the most efficient one which will be used in the experiment.

The covariance of the best filter among different moments is usually uncorrelated. Then the covariance is

$$V_j(k) = E[d_k, d_{k-1}^T] \equiv 0, (j \neq 0) \quad (19)$$

Then we can deduce the specific covariance

$$V_j(k) = H_{k+j} A_{k+j-1} (I - L_{k+j-1} H_{k+j-1}) \cdots$$
$$A_{k+1}(I - L_{k+1} H_{k+1}) A_k [C_{k/k-1} H_k^T - L_k V_0(k)] \quad (20)$$
$$, (j \neq 0)$$

where $V_0(k) = H_k C_{k/k-1} H_k^T + R_k$. The condition of Eq.(19) will be evaluated to be true if and only if

$$C_{k/k-1} H_k^T - L_k V_0(k) = 0 \quad (21)$$

It is almost impossible to get the absolute zero in Eq.(21). We define $S_k = C_{k/k-1} H_k^T - L_k V_0(k)$ and $g_k = \sum_{i=1}^{i=m} \sum_{j=1}^{j=n} S_{ij}^2(k)$.

The target currently is turned to minimize $g_k$. The basic idea to minimize $g_k$ is using the optimization algorithm such as steepest descent method. However, it is the iterative algorithm which is not an appropriate method applied under the real condition. Here we assume $Q_k$, $R_k$ and $V_0$ are positive definite. The key of the one-step AKAF algorithm is to deduce the fading factor with one simple step. The optimal forgetting factor can be computed by

$$\lambda_k = \max\{1, \frac{1}{n} tr[N_k M_k^{-1}]\} \quad (22)$$

where $M_k = H_k A_{k-1} C_{k-1} A_{k-1}^T H_k^T$, $N_k = V_0(k) - H_k Q H_k^T - R$ and $tr[]$ is the symbol of the matrix trace. In this algorithm, we still need to calculate the inverse of the matrix $M_k$. This algorithm can be still further simplified and then we get

$$\lambda_k = \max\{1, \alpha \cdot tr[N_k] / tr[M_k]\} \quad (23)$$

Here, $\alpha$ is the scale factor which is adjusted upon the practical condition. The proof of the algorithm above can be found in Ref.[35].

The formula Eq.(23) above is the basic equation which is used in the experiment. When the optimal ratio $\lambda_k$ is obtained, then the optimal gain can be calculated via the followed equations. Currently, the optimal estimation has to be executed step by step.

The gain of the Modified LQG algorithm is

$$L = C_{k/k-1} H^T [HC_{k/k-1} H^T + R]^{-1} \quad (24)$$
$$C_{k/k-1} = \lambda_k A C_{k-1} A^T + Q \quad (25)$$
$$C_k = [I - L_k H] C_{k/k-1} \quad (26)$$

Then the updated gain can be calculated by Eq.(24)-(26). With the extra fading factor, the divergence of the state evolution of LQG method is suppressed.

### D. The modified LQG algorithm

In order to reduce the divergence of the LQG resulted from the different situations, it is possible to combine the methods in section 3.B and 3.C. The first method is to suppress the divergence due to the wrong measured state and the second is to further modify the prediction to achieve the higher precision of the correction. The process is presented in the following.

Step 1: Check the deviation of predicted state from measurement with method in section 3.A to delete the singular value.
Step 2: Estimate the measured noise through Eq.(16) – Eq.(18) and then get the modified gain $L_k$.
Step 3: calculate optimal ration $\lambda_k$ with method in Section 3.C.
Step 4: calculate the predicted state via Eq.(8) and control vector $u_k$ via Eq.(12).
Step 5: repeat step 1 to step 4 to correct the continuous wavefront.

Since the modified method should be executed on the real time, it still increases the computation burden. The most consuming time of the calculations is the matrix operation, so there is superiority to test it based on Matlab Software from Mathworks Company in the section 4.

### 4: Laboratory study on MLQG

### A. The model of the laboratory study

The model of the experiment is the classical Adaptive Optics (AO) model as showed in Fig. 1. We calculate the gain on each step rather than the fixed gain calculated offline before the execution as described in section 3. In our experiment, the basic calculation model is listed below.

The effective aperture pupil $D$ of setup is 4.5 mm. The response time of DM is 1ms. The sampling frequency of WFS is 407 Hz, which is composed of 32*32 lenslet array and a 2048*1088 pixel camera. This is an approximated two-frame-delay system depending on our estimation. The input states are the Zernike modes of the wavefront. Since the actuator number of our DM is 52, the first 52 zernike modes are selected to be corrected and totally measured mode number is 150. The tip/tilt stroke of DM is bigger than 50 μm and wavefront inter-actuator stroke is bigger than 3μm. The CPU of the computer is Intel@ core I7-3720. The wavefront is introduced by the phase screen (PS) which is quite rough and not fitted to the Kolmogrov turbulence model very well. The turbulence is produced by PS with $D/r_0$=90 according to the statistical variance of Zernike mode coefficients in Fig. 1(b), where the theoretical value is derived from Noll in

1976 [23]. This will of course result in the severe turbulence and the divergence. The Power Spectrum Density (PSD) of the PS is showed in Fig. 2(a). The abrupt decreasing of the spectrum is usually named by cutoff frequency and is about $f_c=3V/D$ which is about 10 times bigger than the Tyler cutoff frequency and is set by the cutoff frequency of the model. After the cutoff frequency, the spectrum decreases with $f^{-5/3}$ and then $f^{-7/3}$ law, not followed by the Kolmogorove spectrum $f^{-11/3}$ law continuously [1]. This fits the assumed turbulence model in section 2, in which the higher frequency error takes the bigger proportion.

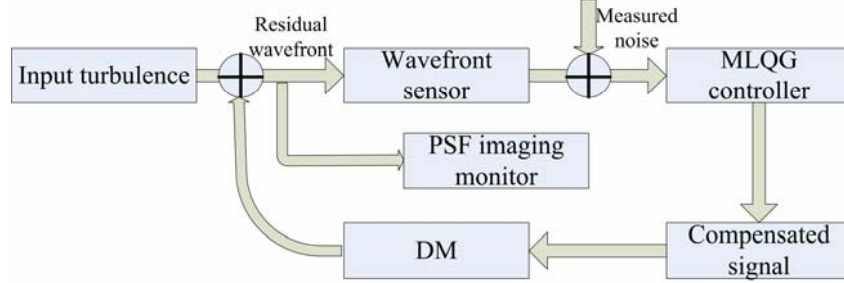

Fig. 1. Schematic diagram of the close-loop model of Adaptive Optics.

The basic Proportional-Integral (PI) controller is tested in our system as a reference in our new method. The updated state equation is discretized as Eq.(5). The most important factor is $G_n$ which determines the optimal control. There is also a series of the predicted control algorithm such as Least Square (LS) predicted algorithm to optimize $G_n$. However, these methods are not optimal. For the simplification here, the gain is set by a fixed constant 0.5. The different gains are also tested, but there is no apparent improvement.

**B. Checking the control divergence**
Regarding to the setup discussed in this paper, the normal LQG control is not able to be applied successfully since the control diverges after several minutes when the loop is closed to correct the error introduced by the PS. Therefore, the different aspects are checked. Firstly, the calculation error of the third reason in the subsection 3.A is ignored thanks to the default data on the kind of double-precision floating point in MATLAB software. The tip and tilt are a little bit smaller than the Kolmogorov statistic. Beyond that, it also denotes that the first and the second aspects in subsection 3.A may appear in the close-loop control. Fig. 3 shows that both reasons appear in the practical control.

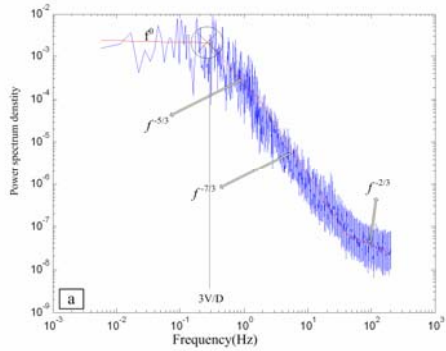 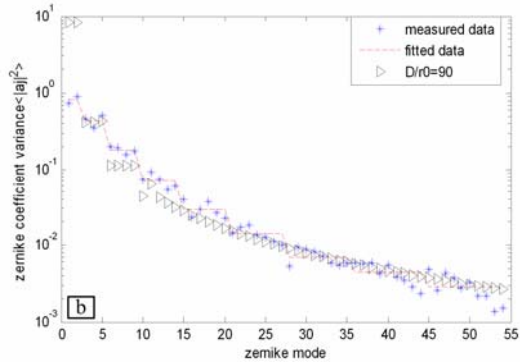

Fig. 2. Statistical turbulence of the phase screen. (a) PSD of angle of arrival under the wind speed 1°/s. (b) Variance of Zernike coefficients of wavefront. The triangle line represents the theoretical values.

Fig. 3a shows the variation of $\lambda$ and Fig. 3b presents the singular values of the gain $L_k$ in the circles. Fig. 4 also shows that the abrupt noise and the control diverging appear on the correction with LQG method. These indirectly state that the LQG method is fragile to the strong turbulence.

In addition, there is another defect of the setup, that the imaging camera is not accurate enough. With the same laser power, since the PS dims the PSF severely, the image pixels on camera will be saturated on the closed loop if the image on open loop can be detected. To the contrary, the image on open loop will be difficult to be recorded if the image pixels are not saturated on closed loop, while you can only record the random noise. Therefore, the PSF on both open loop and closed loop can be only recorded on different laser power levels or the different integration time of camera.

**C. Experimental results**

**C.1 Monitor the MLQG control method**
The gain factor $L_n$ and scale $\lambda$ are the main monitoring objects. They will vary when the turbulence produced by the PS changes violently. If there is the sudden noise similar to unexpected impulse, the evolution will be very likely to diverge. The shortcoming is that the specific modes related to the noise will be ignored and thus the residual error will be increased. The gain $L_n$ will be amplified by multiplying $\lambda$ when the variance of the wavefront increases. In this sense, the modified method is much more like a kind of filter of the LQG.

The variation of scale factor $\lambda$ is plotted in Fig. 3(a). After about 6.5s, it starts to fluctuate. This states that the wavefront variation goes beyond the initially statistics. There are total 52 modes to be optimized, so the 52 diagonal elements of $L_k$ are plotted in Fig. 3(b), where the crosses dots in the circles are the modes deleted according to the criteria Eq.(18). The elements of $L_k$ also increase gradually along with time, stating the same results as in Fig. 3(a). Apparently, the MLQG starts to work in this situation.

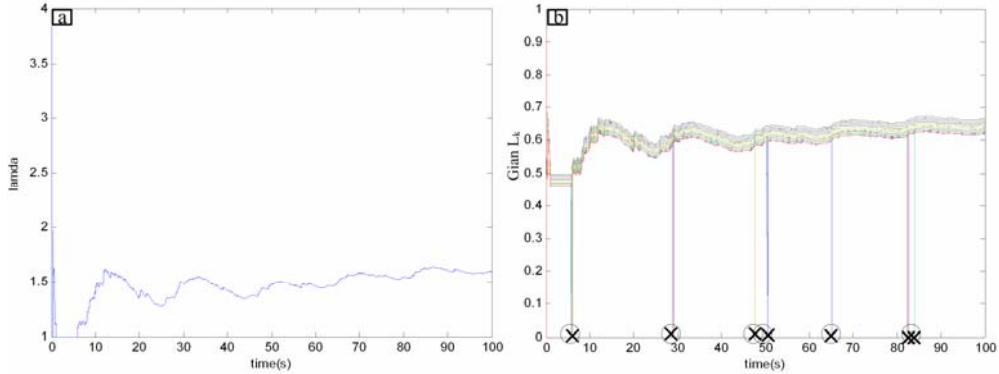

Fig. 3. Parameters variation in the closed loop on 5°/s of the PS. (a) $\lambda$ variation with the time (b) Gain $L_k$ variation with the time. The black crosses in the circle are the filtered states when the abrupt noise happens.

**C.2 Comparing to the PI control method**

For the convenient comparing, the RMS on closed loops with PI, LQG and MLQG and on open loop is plotted in Fig. 4. The RMS on open loop varies dramatically, even achieving over 4 um at a certain moment the 130[th] second when the error is still in the correction range of the DM. The residual RMS of wavefront after correction by PI control is a little bit bigger usually than that with MLQG method. The residual RMS after correction by PI is about 1/5 $\lambda$ while that is about 1/6 $\lambda$ with MLQG method.

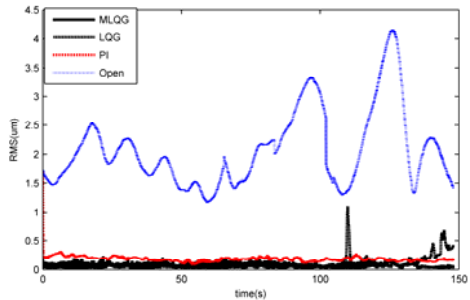

Fig. 4. RMS of the residual errors compares to that on open loop when the distorted wavefront is corrected by different methods including MLQG, LQG and PI at 1°/s of rotation speed of PS.

In the following, LQG method is not considered any more with which the correction usually diverges. The advantage is quite apparent for the specific Zernike modes, the first, the third, the 5[th] and the 52[th] order in Fig. 5. This also states the higher bandwidths of the closed loop where the cross point of spectrums of both open loop and closed loop stands for the effective bandwidths. The other modes are similar to the modes in Fig. 5. Both Fig. 4 and Fig. 5 present the aberration of low temporal frequency taking the main part in the wavefront error.

Considering the performance of the set up, the Full Width Half Maximum (FWHM) is also recorded in Fig. 6. Referred to the reason above, the image intensity is normalized. The FWHM of the open-loop PSF is 142 pixels which are about 10 times larger than that on close loop with different methods. The FWHM of long-exposure PSF under the different rotating speed (1°/s, 3°/s, 5°/s) of PS is also 16 pixels, 18 pixels and 19 pixels respectively when the loop is closed with MLQG, while it is 12 pixels without PS. This shows that the correction doesn't achieve the designed goal. It is because that DM is only composed of 52 actuators and thus at most 52 modes can be corrected, while the higher modes take a big part in the wavefront aberration. According to the statistical estimation of the PS, the first 52 orders take up about 99% of the wavefront and the residual errors after correction with both PI and MLQG methods take about 47% of the residual errors when the wavefront is decomposed by 150 modes. In all recorded data, the modes coefficients after 52ed mode (amplitude varies in the range of 10 nm) are comparable with the first 52ed (amplitude varies in the range of 10-20 nm) and even worse in a few cases. This partially explain that why the correction of turbulence can't achieve the best one without PS.

The residual RMS also shows the faster of the rotating PS, the worse of the performance. The Strehl ration of close loop with MLQG is 9.56 percent higher than that with PI control way and about 40 times higher than that on the open loop.

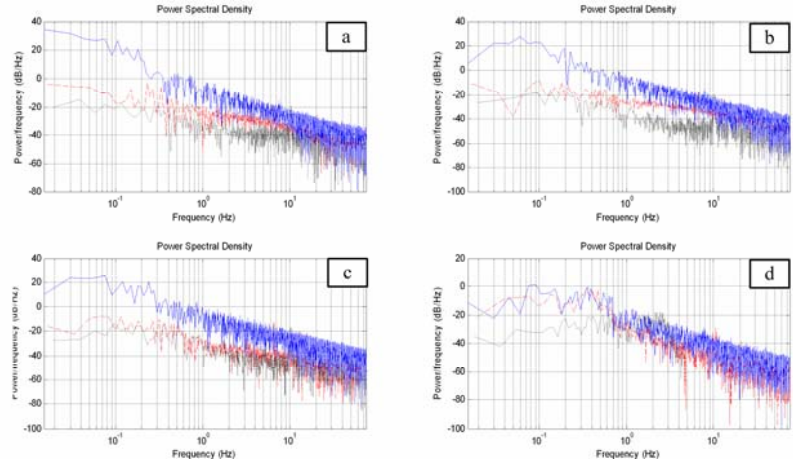

Fig. 5. Close-loop PSD with different algorithms. Open loop is on the blue line. Close loop with PI algorithm is on the red-long-dash line. Close loop with MLQG algorithm is on the black-short-dash line. (a) The PSD of the 1ed Zernike mode (tip). (b) The 3ed Zernike mode (defocusing). (c) The 5ed Zernike mode. (d) The 52ed (the last mode) Zernike mode.

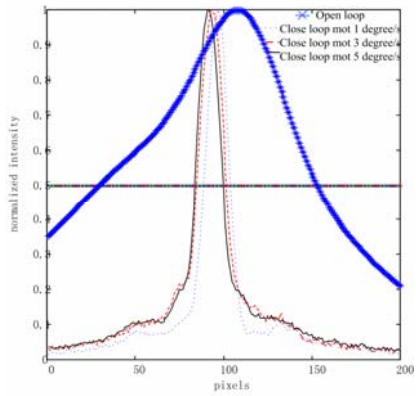

Fig. 6. Close-loop PSF after correction with LQG controlling method at the different wind speeds (1º/s, 3º/s, 5º/s). The intensity is normalized for all images.

The power spectrums of both open loop and closed loop are plotted in Fig. 7. The comparing between open loop and close loop shows the superiority of MLQG to the normal PI method. We sample all the data on one circle of PS and compare the partial data of the sampling to the whole spectrum, getting the same results as in Fig. 7. Since the control software is Matlab from MathsWorks Company, it is impossible to get the very high bandwidth. Therefore, the bandwidth in Fig. 7 is around 2-5 Hz. The LQG controlling can still improve it comparing to PI controlling in Fig. 7. Fig. 8 also shows the better long-exposure PSF of the MLQG controlling than that with PI controlling when the PS is rotated at 1º/s. The second ring of the diffraction is clear on the long-exposure PSF with the controlling of MLQG. The Strehl after correction with MLQG is 0.45 and that with PI is 0.36; the bandwidths are 5 Hz and 2.1 Hz respectively.

This agrees with the residual RMS in Fig. 4 according to the estimation $Strehl = \exp(-\sigma^2)$. We didn't observe the apparently longer delay of MLQG compared to PI controlling, where the consumption of calculation time of the former is prolonged about 1/3 sampling period than the latter. This short latency is ignored in the model and the dynamic characteristic of DM is also neglected.

**D. Discussion**
In addition to the advantages of mentioned above, the shortcomings should be stressed too. Here, the static error of the setup is ignored since it has been calibrated very well. The Strehl ratio of the setup without PS is about 0.966 and the RMS of the static error is about $\lambda/32$, while the residual RMS on the closed loop with PS is $\lambda/7$ for MLQG and $\lambda/6$ for PI. If static errors of the system cannot be ignored such as the results in the lower Strehl (smaller than 0.9), the static error should also be added into the state equation Eq.(8) as presented in Ref.7.

The first modified method in subsection 3.B is the basic filter to get rid of the singular error. It is very important since sometimes the abrupt noise may lead to the control output out of the scale of a corrector, even worse, damaging the corrector. Definitely, the second one in subsection 3.B dealing with gradually changing errors also plays the same role, with a slight difference in increasing the ratio of adjacent states in the gain to converge the control. Actually, if it is unnecessary to achieve very high precision (may be worse than PI since part of states is discarded), the first one is robust enough to suppress the divergence and can also save the consuming time since the gain $L$ in Eq.(9) is the one and only parameter to be modified and can be calculated off line. Moreover, if the wavefront error is stable and only a few singular values appear, the gain $L$ can be also obtained off-line and modified online according to the rules in subsection 3.B. However, the reason why the

LQG is introduced in most cases is that it can lead to higher performance than the normal control. Therefore, the both modification methods are considered.

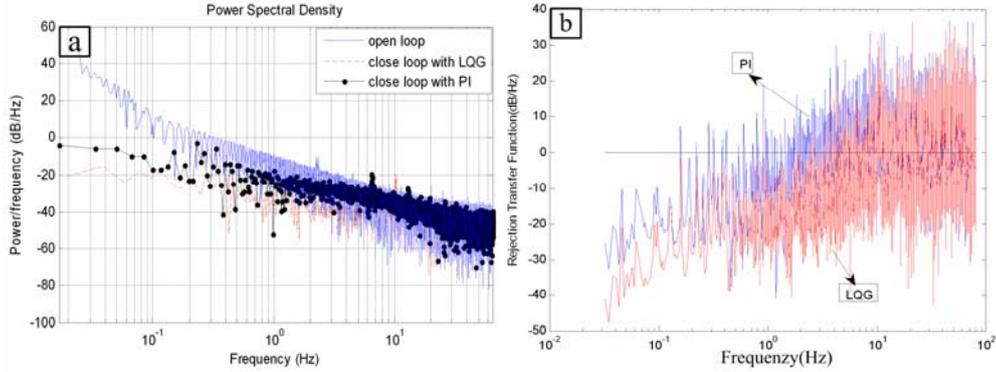

Fig. 7. The power spectrum Comparison of open loop and closed loop with PI and LQG methods at the rotation speed at 1°/s.(a) the power spectrums of open loop and closed loop with LQG and PI methods (b) the rejection function of PI control and LQG control

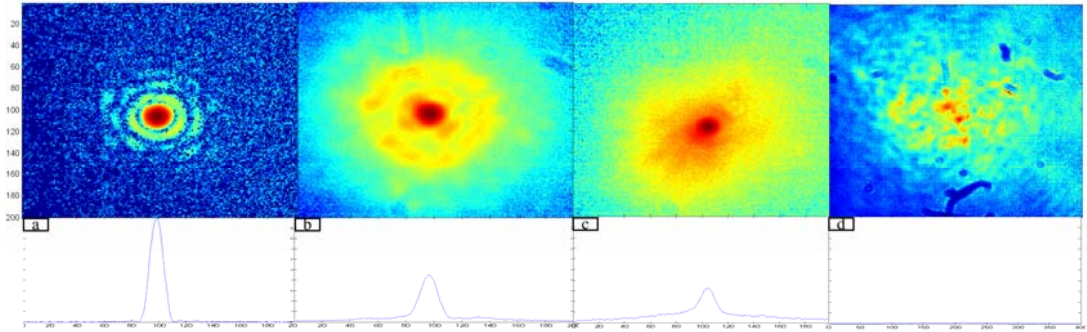

Fig. 8. Long-exposure PSF images. The first three maps consist of 200×200 pixels. The last one consist of 400×400 pixels (a) open loop without PS.(b) close loop with LQG method when PS speed is at 1°/s (c) close loop with PI method when PS speed is at 1°/s (d) open loop when PS is at 1°/s.

In addition, thanks to the strong capability of dealing with the matrix operation, there is no obvious demerit compared to the simple PI control. If it is transplanted to another control software or system, the time consumption of the calculation should be noted.

It is also noted that the modification for all the modes of a wavefront is identical. If $\lambda_k$ was adjusted for each component of gain $L_k$ respectively, the accuracy of correction could be further improved.

Finally, considering that only 52 modes are corrected in the experiment, it is foreseeable that higher performance will be achieved if a better DM with larger number of actuators is selected.

## 5. Conclusion

The LQG algorithm is developed in this paper. Two restrictions have been proposed to modify the LQG algorithm which diverges in our experiment. Given different wind speeds, the Modified LQG can effectively correct the wavefront error while the LQG diverges in most cases. Compared to the original LQG controlling, the gain of the filter in MLQG has to be updated in real time. However, only in this way, the divergence can be suppressed and the controlling can keep stable in our case.

The experiment confirms a successful application of MLQG on AO to correct the dynamically varying turbulence and to deal with abrupt noises. Then compared to the classical control, the correction with MLQG leads to higher strehl ratio according to our test on the strong turbulence. These two modifications can strengthen the LQG controlling potentially when it is applied in practice such as vibration filtering or some other controllings under extreme circumstances.

The author appreciates the great help of Mr. Stefan Hippler and Mr. Pengqian Yang optimizing the setup. The study is partially supported by the MPG-CAS graduate studentship and the German Federal Ministry of Education and Research (BMBF) under grant 05A11VS1.

## References

1. J.-M. Conan, G. Rousset, P.-Y. Madec. "Wave-front temporal spectra in high-resolution imaging through turbulence," Journal of the Optical Society of America A 12, 1559–1570 (1995).
2. Correia, C., Conan, J.-M., Kulcsár, C., Raynaud, H.-F., & Petit, C. "Adapting optimal LQG methods to ELT-sized AO systems," In AO4ELT. Paris.

(2009).

3. C. Correia, H.-F. Raynaud, C. Kulcsár, J.-M. Conan. "Globally optimal minimum-variance control in adaptive optical systems with mirror dynamics," N. Hubin, C.E. Max, P.L. Wizinowich (Eds.), Adaptive optics systems: real time control and algorithms, Proc. soc. photo-opt. instrum. 7015, 70151F (2008).

4. Correia, H.-F. Raynaud, C. Kulcsár, J.-M. Conan. "Minimum-variance control for woofer–tweeter systems in adaptive optics," Journal of the Optical Society of America A 27, A133–A144 (2010).

5. C. Correia, H.-F. Raynaud, C. Kulcsár, J.-M. Conan. "Minimum-variance control for astronomical adaptive optics with resonant deformable mirrors," European Journal of Control 17, 222–236 (2011).

6. C. Correia, J.-P. Véran, L. Poyneer. "Gemini planet imager minimum-variance tip-tilt controllers," Adaptive optics: methods, analysis and applications (AO), Proc. opt. soc. am., OSA, Toronto, Canada (2011).

7. C. Dessenne, P.-Y. Madec, G. Rousset. "Optimization of a predictive controller for closed-loop adaptive optics," Applied Optimization 37, 4623–4633 (1998).

8. R. Fraanje, J. Rice, M. Verhaegen, N. Doelman. "Fast reconstruction and prediction of frozen flow turbulence based on structured Kalman filtering," Journal of the Optical Society of America A 27, A235–A245(2010).

9. D. Gavel, D. Wiberg. "Toward strehl-optimizing adaptive optics controllers," Wizinowich (Ed.), Adaptive optical systems technologies II, Proc. soc. photo-opt. instrum. eng., Vol. 4839 , 890–901 SPIE (2002).

10. E. Gendron, P. Lena. "Astronomical adaptive optics I. Modal control optimization," Astronomy and Astrophysics 291, 337–347 (1994).

11. E. Gendron, P. Lena. "Astronomical adaptive optics II. Experimental results of an optimized modal control," Astronomy and Astrophysics 111, 153–167 (1994).

12. Gilles, L., & Ellerbroek, B. "Computationally efficient, practical implementation of tomographic minimum variance wavefront control using laser and natural guide stars for MCAO and MOAO," In ECC'09. European control conference (2009).

13. J.W. Hardy. Adaptive optics for astronomical telescopes. Oxford, New York (1998).

14. K. Hinnen, M. Verhaegen, N. Doelman. "A data-driven H2-optimal control approach for adaptive optics," IEEE Transactions on Control Systems Technology 16, 381–395(2008).

15. C. Kulcsár, H.-F. Raynaud, C. Petit, J.-M. Conan, B. Le Roux. "Optimality, observers and controllers in adaptive optics," Adaptive optics: signal recovery and synthesis topical meetings, Proc. opt. soc. am., Vol. PMA1, OSA, Charlotte, AWC1, USA (2005).

16. C. Kulcsár, H.-F. Raynaud, C. Petit, J.-M. Conan, P. Viaris de Lesegno. "Optimal control, observers and integrators in adaptive optics," Optics Express 14, 7463–8012 (2006).

17. B. Le Roux, J.-M. Conan, C. Kulcsár, H.-F. Raynaud, L.M. Mugnier, T. Fusco. "Optimal control law for classical and multiconjugate adaptive optics," Journal of the Optical Society of America A 21, 1261–1276 (2004).

18. D.P. Looze. "Minimum variance control structure for adaptive optics systems," Journal of the Optical Society of America A 23, 603–612(2006).

19. D.P. Looze. "Linear-quadratic-Gaussian control for adaptive optics systems using a hybrid model," Journal of the Optical Society of America A 26, 1–9 (2009).

20. D.P. Looze. "Structure of LQG controllers based on a hybrid adaptive optics system model," European Journal of Control 17, 237–248 (2011).

21. S. Meimon, C. Petit, T. Fusco, C. Kulcsár. "Tip-tilt disturbance model identification for Kalman-based control scheme: application to XAO and ELT systems," Journal of the Optical Society of America A 27, A122–A132 (2010).

22. B. Neichel, F. Rigaut, A. Guesalaga, I. Rodriguez, D. Guzman. "Kalman and H-infinity controllers for GeMS. Adaptive optics: methods, analysis and applications (AO)," Proc. opt. soc. am., OSA, Toronto, Canada, JWA32 (2011).

23. R.J. Noll. "Zernike polynomials and atmospheric turbulence," Journal of the Optical Society of America 66, 207–211(1976).

24. R.N. Paschall, D.J. Anderson. "Linear quadratic Gaussian control of a deformable mirror adaptive optics system with time-delayed measurements," Journal of Applied Optics 32, 6347–6358(1993).

25. C. Petit, J.-M. Conan, C. Kulcsár, H.-F. Raynaud, T. Fusco. "First laboratory validation of vibration filtering with LQG control law for adaptive optics," Optics Express 16, 87–97 (2008).

26. C. Petit, J.-M. Conan, C. Kulcsár, H.-F. Raynaud, T. Fusco, J. Montri et al. "Optimal control for multi-conjugate adaptive optics," Comptes Rendus de l'Académie des Sciences, Physique 6, 1059–1069(2005).

27. P. Piatrou, M.C. Roggemann. "Performance study of Kalman filter controller for multiconjugate adaptive optics," Journal of Applied Optics, 46, 1446–1455(2007).

28. L.A. Poyneer, B.A. Macintosh. "Spatially filtered wave-front sensor for high-order adaptive optics.," Journal of the Optical Society of America A 21, 810–819 (2004).

29. L.A. Poyneer, J.-P. Véran. "Predictive wavefront control for adaptive optics with arbitrary control loop delays," Journal of the Optical Society of America A 25, 1486–1496 (2008).

30. L.A. Poyneer, J.-P. Véran. "Kalman filtering to suppress spurious signals in adaptive optics

control," Journal of the Optical Society of America A 27, A223–A234 (2010).
31. T. Fusco, C. Petit, G. Rousset, J.-M. Conan, J.-L. Beuzit. "Closed-loop experimental validation of the spatially filtered Shack–Hartmann concept," Journal of the Optical Society of America A 30, 1255–1257(2005).
32. T. Fusco, G. Rousset, J.-L. Beuzit, D. Mouillet, K. Dohlen, R. Conan, C. Petit, G. Montagnier. "Conceptual design of an extreme AO dedicated to extra-solar planet detection by the VLT-planet finder instrument," Astronomical adaptive optics systems and applications II, Proc. soc. photo-opt. instrum. eng., Vol. 5903, SPIE, San Diego, USA 59030K1–59030K12 (2005).
33. T. Fusco, G. Rousset, J.-F. Sauvage, C. Petit, J.-L. Beuzit, K. Dohlen et al. High-order adaptive optics requirements for direct detection of extrasolar planets: application to the SPHERE instrument. Optics Express 14, 7515–7534(2006).
34. C. Petit, J.-M. Conan, C. Kulcsár, H.-F. Raynaud, T. Fusco, J. Montri, F. Chemla, D. Rabaud "Off-axis adaptive optics with optimal acontrol: Experimental and numerical validation," Proc. SPIE 5903, 227-235 (2005).
35. Xia, Q., Rao, M., Ying, Y., & Shen, X.. "Adaptive fading Kalman filter with an application," Automatica 30, 1333-1338 (1994).
36. Hide, Christopher, Terry Moore, and Martin Smith. "Adaptive Kalman filtering for low-cost INS/GPS," Journal of navigation 56.1, 143-152 (2003).
37. Kim, Pyeongjun, Joohyun An, and Kwanho You. "A Fuzzy Adaptive Fading Kalman Filter Approach for Accuracy Improvement of a Laser Interferometer," Computer Applications for Security, Control and System Engineering. Springer Berlin, Heidelberg, 245-253 (2012).